# Tables of equation-of-state, thermodynamic properties, and shock Hugoniot for hot dense fluid deuterium

Mofreh R. Zaghloul

Department of Physics, College of Science, United Arab Emirates University, P.O. Box 15551, Al-Ain, UAE.

We present computational results and tables of the equation-of-state, thermodynamic properties, and shock Hugoniot for hot dense fluid deuterium. The present results are generated using a recently developed chemical model that takes into account different high density effects such as Coulomb interactions among charged particles, partial degeneracy, and intensive short range hard core repulsion. Internal partition functions are evaluated in a statistical-mechanically consistent way implementing recent developments in the literature. The shock Hugoniot curve derived from the present tables is in reasonable overall agreement with the Hugoniot derived from the Nova-laser shock wave experiments on liquid deuterium, showing that deuterium has a significantly higher compressibility than predicted by the SESAME tables or by Path Integral Monte Carlo (PIMC) calculations. Computational results are presented as surface plots for the dissociated fraction, degree of ionization, pressure, and specific internal energy for densities ranging from 0.0001 to 40 $g/cm^3$ and temperatures from 2000 to ~ $10^6$ K. Tables for values of the above mentioned quantities in addition to the specific heat at constant pressure, $c_p$, ratio of specific heats, $c_p/c_v$, sound speed and Hugoniot curve (for a specific initial state) are presented for practical use.

## I. INTRODUCTION

Reliable predictions of the equation-of-state (EOS) and thermodynamic properties of hot fluid deuterium in the high-energy-density regimes are of great interest to inertial confinement fusion (ICF) research and astrophysical studies. For example, to compress and heat the ICF fuel, the outer layer of the ICF target is bombarded using high-energy laser beams or ion beams (direct drive) or x-rays emitted from a high-Z enclosure or hohlraum (indirect drive). The heated outer layer ablates and explodes outward, producing a reaction force compressing the remainder of the target accelerating it inwards. This process is designed to create sufficiently powerful shock waves that travel inward through the target to compress and heat the fuel at the center to ignition conditions sufficient for efficient thermonuclear burn. Predicting the performance of the ICF target during such an implosion process depends largely on the compressibility or the EOS of the deuterium or deuterium-tritium fuel material.

The challenges of understanding the deuterium EOS at ultrahigh pressures require the development of theoretical models capable of treating the complexities of strongly coupled systems and considering the involved physics of hot dense matter. Rough approximations of the properties of other hydrogen isotopes may be easily obtained through scaling the results for deuterium. Estimation of the equation-of-state of deuterium, over a wide range of densities and temperatures, is also essential to many other fundamental and applied topics and has been under intensive experimental and theoretical investigation for the past few decades [1-33].

A recently developed model, based on the chemical picture (free energy minimization), is used to generate the present data for the EOS and thermodynamic properties of fluid deuterium over a wide range of densities, $\rho$, (from 0.0001 to 40 g/cm$^3$) and temperatures, $T$, (from 2000 to ~ $10^6$ K) [34]. The model takes into consideration the dissociation equilibrium of diatomic molecules and ionization equilibrium of dissociated species with special interest given to the phenomena of pressure dissociation and pressure ionization. The model also implements a recently proposed formulation for the establishment of statistical-thermodynamically-consistent finite electronic partition functions [35-37] where an occupational probability is prescribed in advance to assure a smooth truncation of the electronic partition function and manifestation of the phenomenon of pressure ionization. A similar formulation and analogous occupational probability is also introduced and used to smoothly truncate the vibrational-rotational molecular partition function of deuterium molecules. To avoid cutting-off the ground states, when using occupational probabilities, which causes a nonphysical vanishing of the partition functions, a reform proposed in Ref. [38] is implemented in the model and applied in the present calculations of electronic and rotational-vibrational molecular partition functions. Electron degeneracy effects together with Coulomb and excluded-volume configurational terms are all taken into account in the model and the resulting free energy function is then used to calculate the occupation numbers and to find the corresponding set of nonideal thermodynamic properties. Model predictions are exceptional in showing the correct physical behavior of non-intersecting isotherms and non-negative slope of isotherms of the packing fraction as function of density. Details of the model, set of governing equations and the used computational scheme can be found elsewhere [34].

## II. MODEL VALIDATION

Determining whether an EOS table is valid for a given regime is a multistage procedure that embodies a number of requirements such as obeying thermodynamic consistency, being free of major computational flaws, in addition to fair or acceptable agreement with existing trustworthy experimental data. Thermodynamic consistency of the present model predictions is assured through using a single free energy function over the whole considered domain for the calculation of occupation numbers and thermodynamic functions. On the other hand, one may consider model predictions free of major computational flaws if they accurately produce the well-known and well-understood limiting physical behavior of the properties in addition to, at least, order of magnitude agreement with predictions from other widely accepted models. In the following subsection (subsection A) we discuss the limiting behavior of the results of the present model and compare model predictions to predictions from first principle Path Integral Monte Carlo (PIMC) simulations. In subsection B, we compare analytical predictions of the Hugoniot curve from the present model to values derived from experimental measurements along with values from other EOS models and tables.

### A- Limiting behavior and comparison with PIMC results

Consistent computations of dissociation and ionization equilibria are involved even for the simplest case of hydrogen isotopes. Consequently, a necessity arises to ascertain confidence that code predictions are free of major computational flaws and that the accuracy of the results is determined exclusively by its underlying physics. For this sake we examine the limiting physical behavior of the computed pressure and internal energy in the low-density, high-temperature region where deuterium is known to behave like an ideal gas of fully-dissociated, fully-ionized species. Concurrently, we compare results from the present model to recent results from first principles PIMC simulations of the deuterium equation-of-state [31]. From among different first-principles Quantum Monte Carlo (QMC) methods and simulations (for a review of different approaches, see [39]), the PIMC results are used in this comparison because of the availability of these results in a tabular form [31]. However, an overall discussion including other QMC methods will follow this relatively detailed comparison with the PIMC results.

Figure 1 shows a comparison between the pressure calculated from the present model and the pressure calculated from PIMC simulations for *a*- low temperature isotherms, and *b*- high temperature isotherms. It has to be noted that the tables of the PIMC results [31] include some anomalous points that we did not include in the figure (e.g. at $T$=62500 K, $\rho$= 0.0841898 g/cm$^3$ and at $T$=95250 K, $\rho$= 0.199561 g/cm$^3$). A similar comparison for the internal energy is shown in both parts of Fig. 2. It is clear from both parts of Fig. 1 that pressure isotherms predicted from the present model show the linear behavior characteristic to ideal gases at the limiting low-density region. It is also recognizable that the values predicted from the present model in the high-temperature, low-density region meet those values predicted from the ideal gas EOS in that region of the $\rho$-$T$ plane. On the logarithmic scale, the pressure of an ideal gas is given by

$$\log_{10}(P) \approx 3.92 + \log_{10}(T) + \log_{10}(\rho) \tag{1}$$

where $P$ is the absolute pressure, $T$ is the absolute temperature and $\rho$ is the mass density. For a temperature of 125000 K and density of 0.1 kg/m$^3$ (10$^{-4}$ g/cm$^3$), one gets $\log_{10}(P) \approx 8$ in agreement with the value shown in Fig. 1-*a*. The corresponding value for the specific internal energy of the fully-dissociated, fully-ionized ideal deuterium gas is the sum of ionization and kinetic energies $\approx 2.2 \times 10^3$ MJ/kg, which also agrees with the value shown in Fig. 2-*a*. It has to be noted that, in accordance to the common practice, energies are referred to a *zero-of-energy* point taken as the energy of dissociated atoms at infinite distance apart in their ground states. Similar checks can be performed for other isotherms in this region of the $\rho$-$T$ plane confirming the convergence of the limiting values of the pressure and internal energy isotherms predicted from the present model to those of a fully-dissociated, fully-ionized ideal deuterium gas. Similar corresponding tests can be performed for other thermodynamic properties given in the tables included in the Appendix section.

The agreement between predictions of the present model and results from PIMC simulation at low-density and/or high-temperature is another indication that our code is free of major computational flaws and that the remarkable discrepancies between both approaches, at low-temperature and high-density, are mainly due to the underlining physics. The fact that the PIMC is an appealing first principles approach led some researchers using chemical models to modify and parameterize their models to better agree with PIMC calculations in the low-temperature high-density regime. However, it is our belief that the two approaches (chemical picture and the PIMC calculations) have to proceed independently for the following reasons: *a*-first-principle PIMC calculations have some anomalous points, as explained above, that may or may not be indicative of some buried inaccuracy sources, *b*- the PIMC calculations do have approximations and are not completely exact, *c*- more importantly, as it can be seen from the Figs 1 and 2, the PIMC method cannot continue to high density at low temperatures; this limitation of the method at high densities and low temperatures is also recognized in Ref. [40] where it is mentioned that as the plasma temperature decreases the Fermi-sign problem in PIMC method prevents the efficient evaluation of degeneracy effects, *d*- although PIMC results agree with experimental measurements of the shock Hugoniot using Z-pinch data [21,41], and converging explosives [42], they disagree with and fail to reproduce the shock Hugoniot derived from the Nova-laser experimental measurements [9,11], which show an increased compressibility of deuterium around 1 Mbar, *e*- the tables generated from the PIMC computations and presented in Ref. [31] are for the pressure and internal energy only; however, data about other thermodynamic properties that depend on the derivatives of these properties (e.g. specific heats and their ratio, sound speed, etc.), whose physical behavior is important for the validation of the results, were not provided, and *f*- the $\rho$-$T$ grid used to generate the PIMC tables in [31] is relatively coarse and it is hard to decide whether isotherms of PIMC results will intersect or not upon using a finer grid (it is to be noted that we excluded anomalous points from the PIMC results in Fig. 1, that show intersection of isotherms and mechanical instability, even for such a coarse grid).

However, it should not be misunderstood from the above discussion that we claim superiority of the chemical models, in general, or the present model and its engendered results, in particular, over first-principles PIMC method. This is not intended or meant here on top of being conceptually hard to justify. Rather, we thoughtfully show that there are a number of reservations that need to be cleared first before considering first-principles PIMC computations as the principal measure, based on which authors may consider modifying their models and approaches to match its predictions. Not the least of these reservations is the failure to continue computations for higher densities at low temperatures (the Fermi-sign problem) and the presence of anomalous points in the resulting tables. The disagreement with the shock Hugoniot derived from the Nova-laser experiments is not a decisive factor yet, pending resolving the puzzle of disagreement between experimental measurements from the Z-machines and converging explosives, on one side, and the Nova-laser on the other side. An overall assessment follows in the next subsection, in which predictions of the principal shock Hugoniot from all models, including other QMC simulations, are compared against experimental measurements.

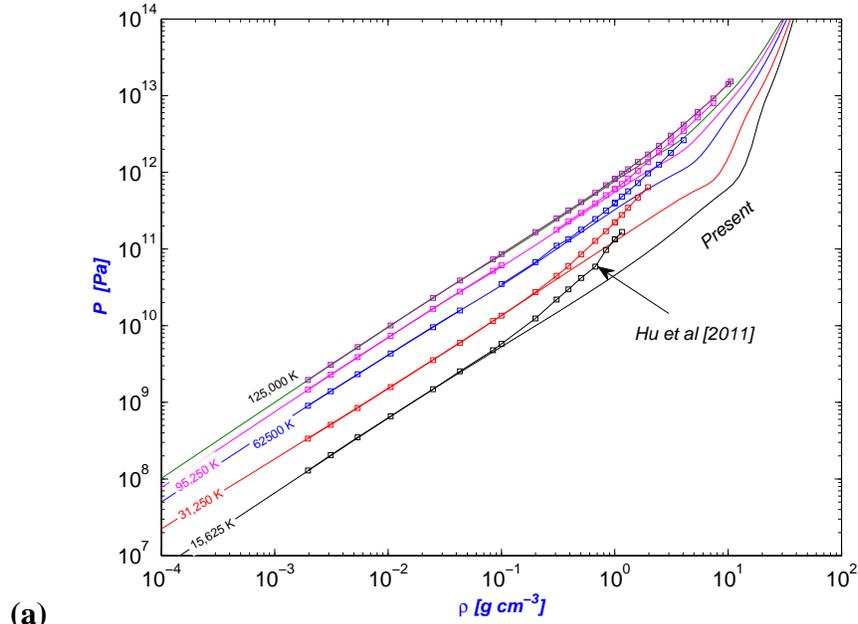

(a)

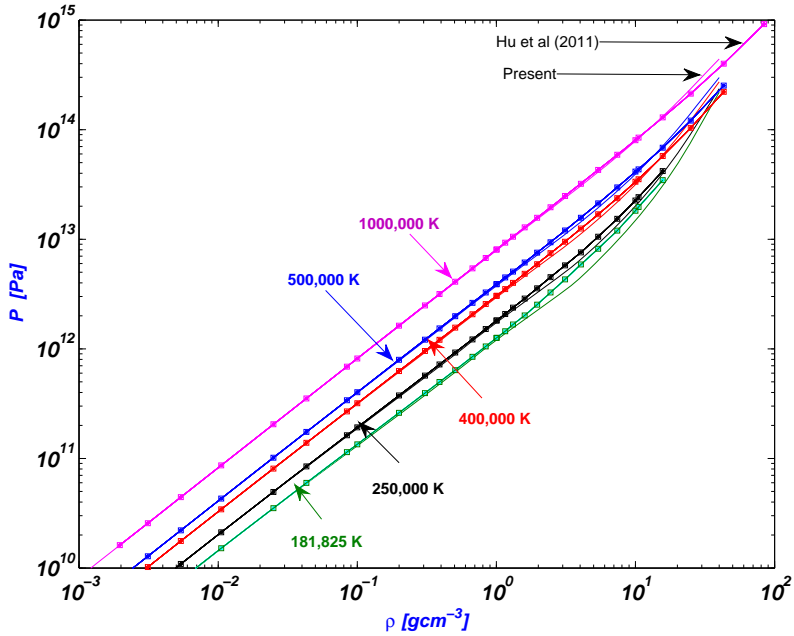

(b)

Figure 1: Comparison between pressure isotherms calculated from the present model (curves without markers) and those calculated from PIMC simulations (curves with markers) for *a*- low temperatures, and *b*- high temperatures.

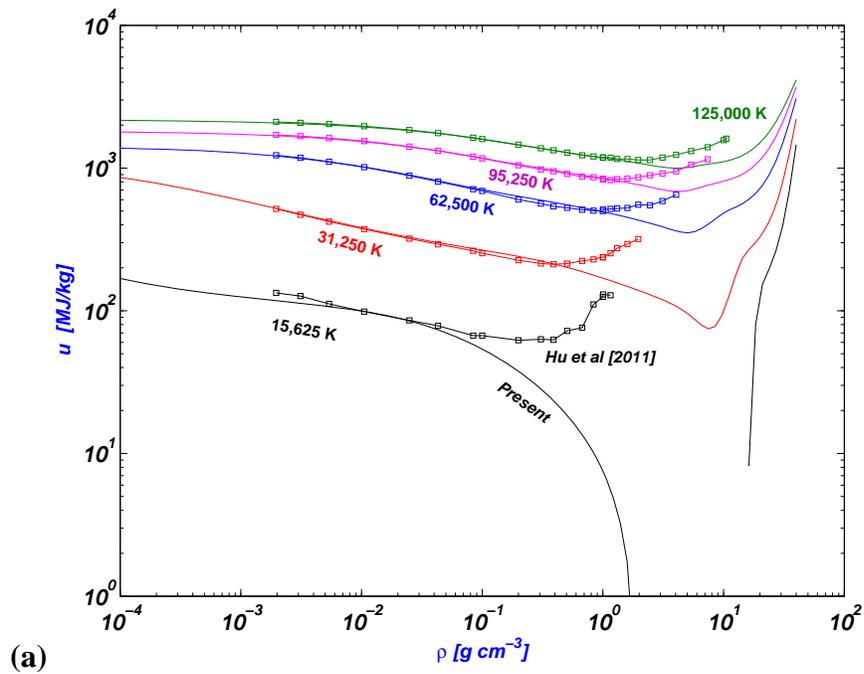

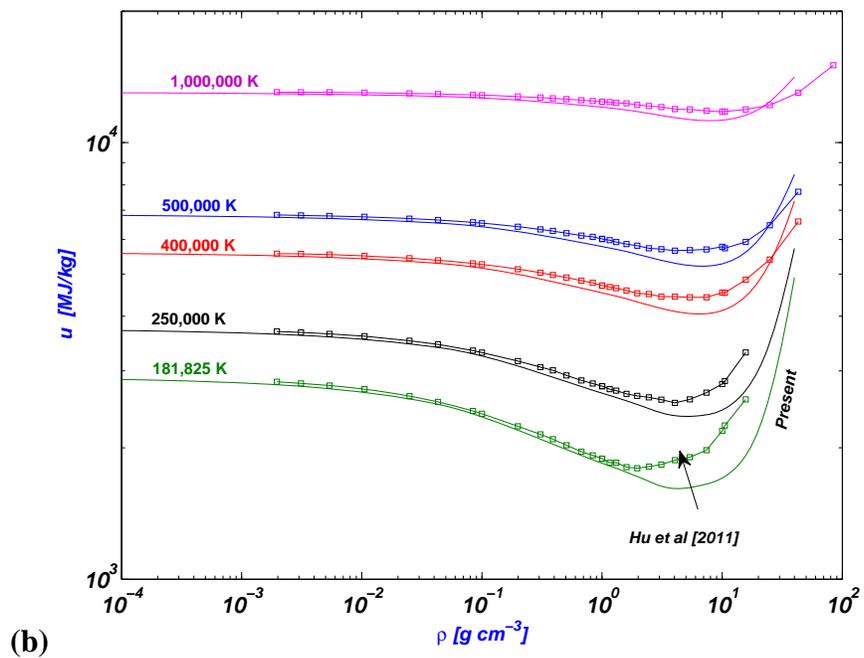

Figure 2: Comparison between isotherms of the specific internal energy calculated from the present model (curves without markers) and those calculated from PIMC simulations (curves with markers) for *a-* low temperatures, and *b-* high temperatures.

**B- Shock Hugoniot**

The shock Hugoniot is the locus of the sequence of thermodynamic equilibrium states reached behind each shock for a sequence of different-strength shocks from a given initial state. The fact that the Hugoniot curve could be determined from experimental measurements provides an important means for benchmarking and validating theoretically generated equations-of-state. The Hugoniot equation can be derived from the mass, momentum and energy balance and can be written as

$$H(\rho_2, P_2) = (u_2 - u_1) - \frac{1}{2}(P_1 + P_2)\left(\frac{1}{\rho_1} - \frac{1}{\rho_2}\right) = 0, \tag{2}$$

where $u$ is the specific internal energy, $P$ is the pressure, $\rho$ is the density and the subscripts 1 and 2 refer to the initial and final states, respectively. Since $P(\rho,T)$ and $u(\rho,T)$ are constrained by a given temperature, the dependent variable becomes the density. Therefore, we can provide the Hugoniot-solving-routine with arrays of density, $\rho$, temperature, $T$, pressure, $P(\rho,T)$, specific internal energy, $u(\rho,T)$, and a temperature grid along which to search for the zero of the function in (2) to a predetermined acceptable tolerance.

Figure 3 shows a comparison of the Hugoniot derived from experimental measurements (laser shock wave experiments [9,11], converging explosive-driven shock waves [42] and Z-pinch-driven compression [41]) with predictions from the present model and several other models including PIMC calculations [17], SESAME model by Kerley [1], linear mixing model by Ross [10], Saumon *et al* [43], Knaup *et al* [44], Filinov *et al* [45], Holst *et al* [46], and Tubman *et al* [47]. Measurements performed in the Nova laser facility where a shock wave is generated by a laser pulse [9,11], show that deuterium has a significantly higher compressibility (maximum compression ~ 6.5) than derived from the standard SESAME equation-of-state and from measurements using hemispherically converging explosives [42] and magnetically-driven flyers using the Z-Machine at Sandia [41] (maximum compression ~ 4). The comparison shows that the results from the present model, from the linear mixing model by Ross and from Saumon *et al* are in reasonable overall agreement with laser shock wave experimental measurements. All three models show increased compressibility (maximum compression ratio around 6) in addition to a behavior close to the behavior of these experimental data though the bump shown from our calculations is at a relatively lower pressure (~0.3 Mbar). Results from PIMC calculations and SESAME model by Kerley are relatively far from these experimental results except for very high pressure where both of the results from the present model and PIMC calculation converge and agree with each other. It has to be noted that at high pressures, the maximum compression ratio for a monatomic ideal gas cannot exceed 4.

Results from other models including first principles approaches vary in value and behavior. The results from wave-packet molecular dynamics (WPMD) performed by Knaup et al [44] showed a principal Hugoniot with a behavior and maximum compression value close to the Nova laser data. However, the accuracy of the WPMD is limited to the semiclassical regime and its applicability at the relevant temperatures was questioned [39]. The results from direct path-integral Monte Carlo (DPIMC) study performed by Filinov et al [45], showed a maximum compression of ~5 at 1.1 Mbar (lowest pressure shown) while quantum molecular dynamics calculations by Holst et al [46], showed a maximum compression ~4.5 and behavior closer to experimental data from converging explosives [42] and Z-pinch [41].

In a recent work [47], the region of the maximum compression along the principal Hugoniot, where the system undergoes a continuous transition from a molecular fluid to a monatomic fluid, is studied using Coupled Electron-Ion Monte Carlo Simulations (CEIMCS). As described by the authors, the calculations used highly accurate Monte Carlo methods for the electrons and included all relevant physical corrections. A maximum compression ratio of ~4.85 was predicted to occur at ~0.5 Mbar using these simulations. This maximum compression at the molecular dissociation crossover is ~5.5% larger than previous predictions from density functional theory (DFT) and ~15% higher than experimental data from Boriskov et al [42] and Knudsen et al [41] and it was concluded that systematic errors might exist in these recent experimental results with a need to resolve them in future experiments. Although, compared to PIMC results, these recent results from the highly accurate CEIMC simulations are slightly shifted in the direction of the predictions of the present model (in terms of the maximum compression and its corresponding pressure), they, at the same time, show that predictions of the present model disagree not only with PIMC but also with more accurate Quantum Monte Carlo methods that work at low temperature and high density.

The fact that predictions of the present model produce the expected limiting behavior and compare well with laser shock wave experimental measurements and a few other theoretical models validates and gives some degree of creditability to the results obtained from the present model as a reasonable approximation of the deuterium equation-of-state. An exceptional feature of the present results is that they provide a complete set of thermodynamic properties.

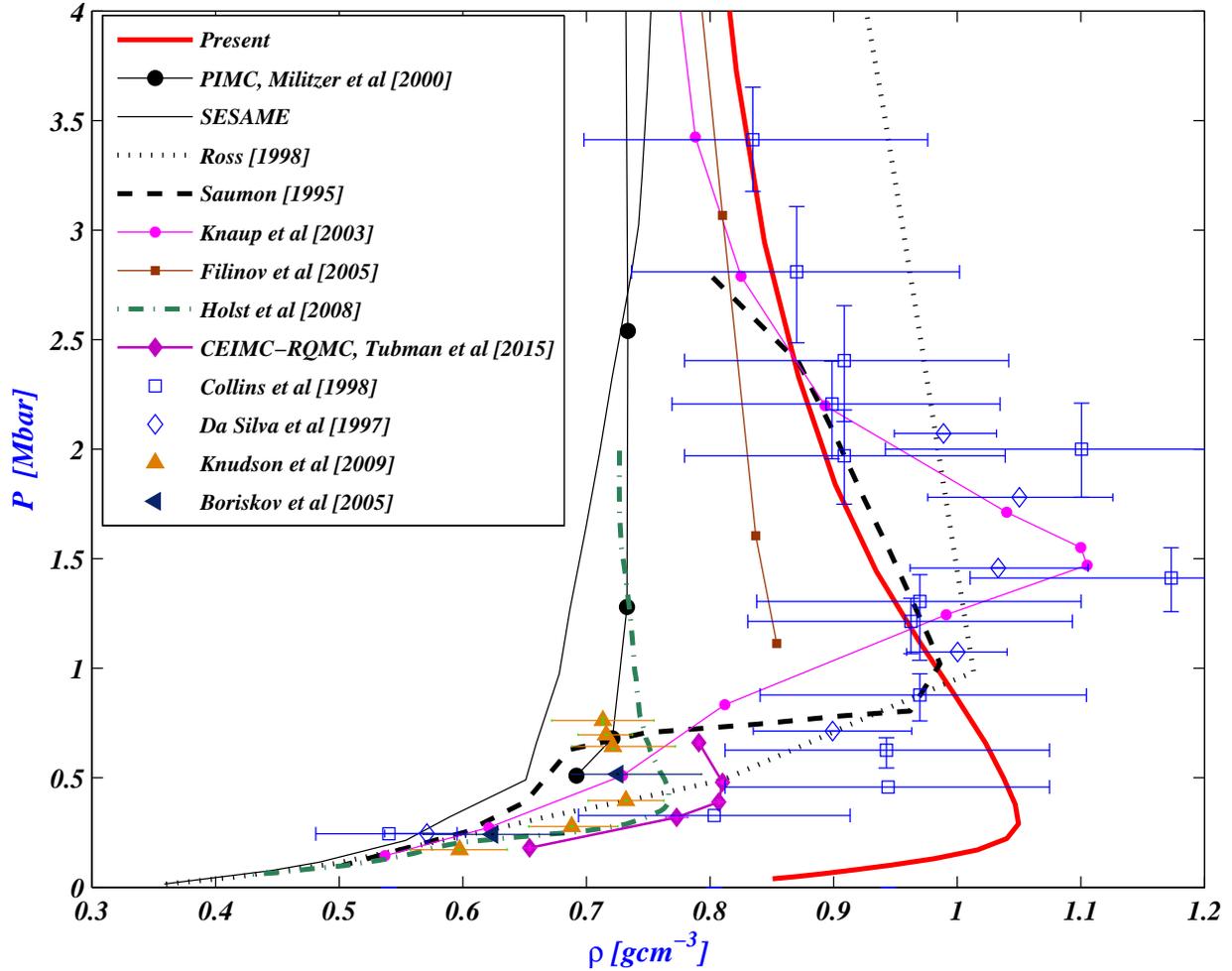

Figure 3: A comparison of the Hugoniot derived from experimental measurements of Da Silva *et al* [9] and Collins *et al* [11], Boriskov *et al* [42] and Knudson *et al* [41] with predictions from the present model and several other models including PIMC calculations [17], SESAME model by Kerley [1], linear mixing model by Ross [10], Saumon *et al* [43], Knaup *et al* [44], Filinov *et al* [45], Holst *et al* [46], and Tubman *et al* [47].

## III- COMPUTATIONAL RESULTS

A surface plot of the degree of ionization, defined as $(n_{D+}/2n_0)$ where $n_0$ is the original number of molecules is shown in Fig. 4, for a wide range of densities (0.0001 to 40 g/cm$^3$) and temperatures (2000 to $10^6$ K). As it can be seen from the plot, there is an enhancement in ionization at intermediate densities (0.1-10 g/cm$^3$) and relatively low temperatures (<50000 K) due to Coulomb interactions. This enhancement in ionization is suppressed at higher densities by quantum effects (degeneracy of free electrons). At extremely high densities, pressure ionization occurs at all temperatures due to hard core repulsion of extended species where equilibrium occupation numbers are shifted towards smaller volume particles (deuterons).

Similarly, a surface plot for the dissociation fraction, $f=(n_D + n_{D+})/2n_0$, for the same wide range of densities and temperatures, is shown in Fig. 5 where pressure dissociation appears at high densities and all temperatures $\leq$ 50000 K.

Surface plots of the total pressure and of the specific internal energy of deuterium fluid are shown in Figs. 6 and 7, respectively. The pressure appears to behave ideally at low densities and/or high temperatures. However, deviations from this ideal behavior become clear at high densities and relatively low temperatures. The negative values of the specific internal energy at low temperatures are indicative of the presence of a significant amount of dimers.

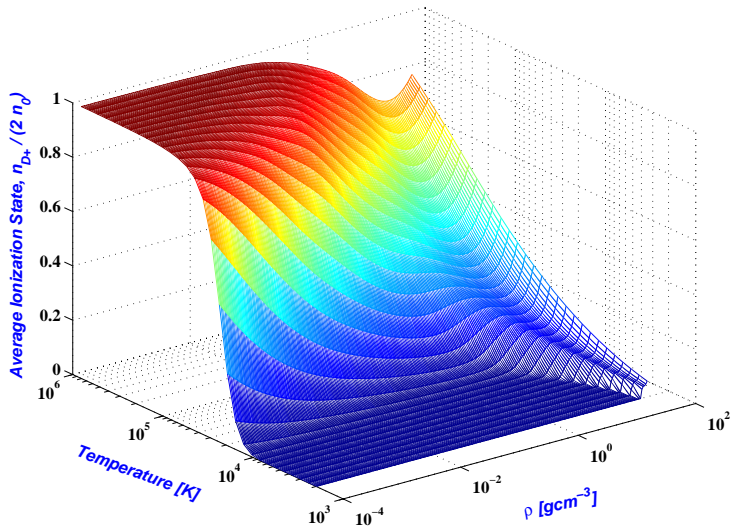

Figure 4: A surface plot of the degree of ionization, $n_{D+}/2n_0$

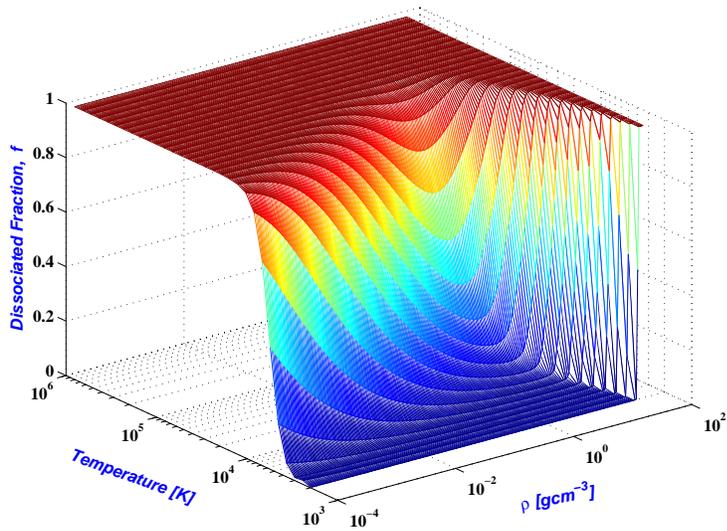

Figure 5: A surface plot of the dissociation fraction, $f = (n_D + n_{D+})/2n_0$

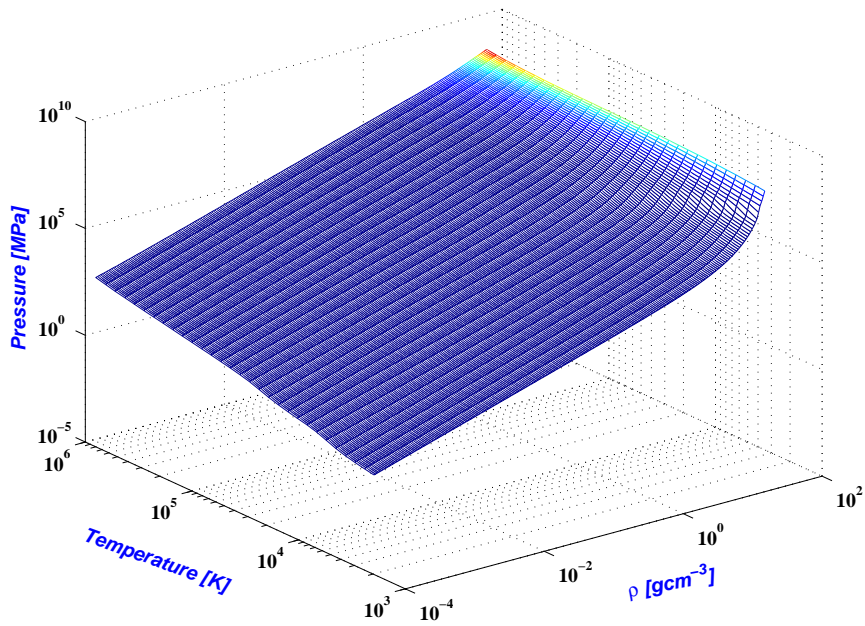

Figure 6: A surface plot of pressure of fluid deuterium

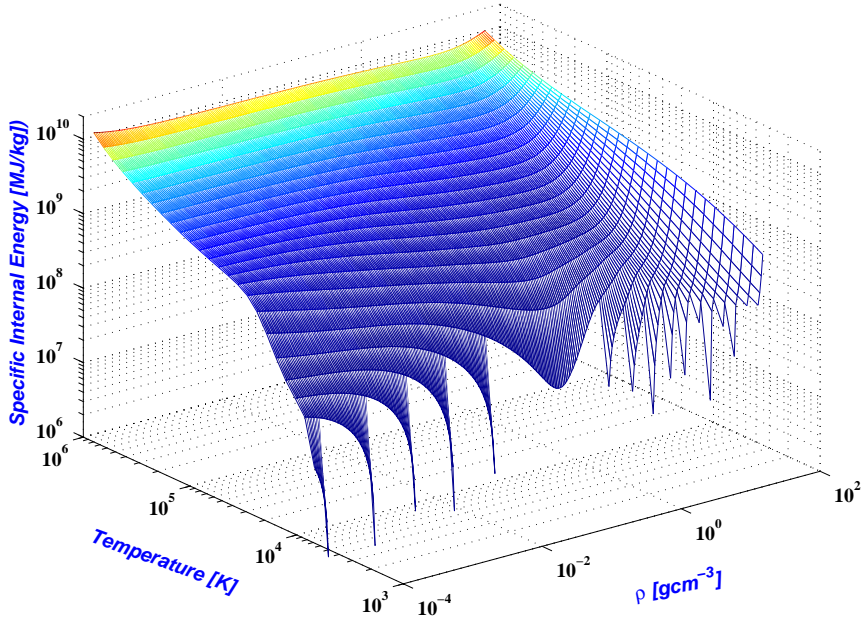

Figure 7: A surface plot of the specific internal energy of fluid deuterium

For practical use, we present a set of tables for values of the EOS and thermodynamic properties of fluid deuterium in the Appendix. In addition to ionization and dissociation fractions, pressure and specific internal energy, tables of specific heat capacity at constant pressure, $c_p$, and ratio of specific heats, $c_p/c_v$ in addition to sound speed are also presented. The data for the shock Hugoniot derived from the present model (for an initial state relevant to experimental measurements) are also given in a separate table.

## IV- CONCLUSIONS

A recently developed consistent chemical model for strongly-coupled deuterium fluid is used to generate tables for the EOS and thermodynamic properties for fluid deuterium over a wide range of densities and temperatures. The shock Hugoniot curve derived from the present tables shows that deuterium has a significantly higher compressibility than predicted by experimental measurements using converging explosives or Z-pinch, though in agreement with predictions from Nova laser shock wave experiments on liquid deuterium. Compared to other theoretical models, the present model is located among a group of models (chemical and first principles QMC) that predict a soft principal Hugoniot while another group of models (SESAME and other first principles QMC) predict a stiff Hugoniot. Tables for the degree of ionization, dissociation fraction, pressure and specific internal energy, in addition to specific heat at constant pressure, ratio of specific heats and the sound speed are generated and presented for practical use. For referencing, the data for the shock Hugoniot derived from the present model are also given in a separate table.

## ACKNOWLEDGMENTS

I would like to acknowledge useful discussion with Burkhard Militzer (University of California, Berkeley) and Suxing Hu (University of Rochester) regarding their work and EOS tables generated using first-principles PIMC calculations. I am also grateful to D. M. Ceperley (University of Illinois, Urbana-Champaign) and C. Pierleoni (University of L'Aquila, Italy) for providing me with results of theoretical models included in Figure 3.

# APPENDIX
# TABLES OF EQUATION-OF-STATE AND THERMODYNAMIC PROPERTIES

All quantities are given by the mantissa and order of magnitude. Thus a tabulated value of temperature written as 2000+04 should be read as $T=0.2000\times10^4$ and a value of density written as 7606-02 should read as $0.7606\times10^{-2}$ and so for all tabulated properties.

Table 1 *Ionization state, $n_{D+}/2n_0$*

| T, K | Density, g/cm³ | | | | | | | | | | | |
|---|---|---|---|---|---|---|---|---|---|---|---|---|
| | 1000-03 | 4237-03 | 1795-02 | 7606-02 | 3223-01 | 1365+00 | 5785+00 | 2451+01 | 1038+02 | 2000+02 | 3000+02 | 4000+02 |
| 2000+04 | 1727-17 | 8396-18 | 4080-18 | 1982-18 | 9630-19 | 4680-19 | 2273-19 | 1104-19 | 5367-20 | 7732-20 | 1510-13 | 4522-01 |
| 2460+04 | 2256-14 | 8084-15 | 2941-15 | 1091-15 | 4155-16 | 1644-16 | 6989-17 | 3680-17 | 6756-17 | 3212-15 | 1072-06 | 5108-01 |
| 3027+04 | 3355-11 | 1247-11 | 4747-12 | 1863-12 | 7619-13 | 3305-13 | 1586-13 | 1003-13 | 2999-13 | 4541-11 | 2105-01 | 5803-01 |
| 3723+04 | 1331-08 | 5210-09 | 2096-09 | 8779-10 | 3892-10 | 1869-10 | 1026-10 | 7950-11 | 4060-10 | 1995-07 | 4101-01 | 6621-01 |
| 4580+04 | 1669-06 | 7116-07 | 3076-07 | 1385-07 | 6666-08 | 3537-08 | 2210-08 | 2073-08 | 1715-07 | 2003-04 | 5806-01 | 7573-01 |
| 5635+04 | 7223-05 | 3568-05 | 1718-05 | 8434-06 | 4403-06 | 2553-06 | 1775-06 | 1936-06 | 2220-05 | 8397-02 | 6479-01 | 8669-01 |
| 6931+04 | 1254-03 | 7218-04 | 4031-04 | 2215-04 | 1261-04 | 7885-05 | 5923-05 | 7083-05 | 9391-04 | 2840-01 | 7124-01 | 9919-01 |
| 8527+04 | 1169-02 | 7085-03 | 4463-03 | 2790-03 | 1750-03 | 1172-03 | 9254-04 | 1147-03 | 1692-02 | 4765-01 | 7894-01 | 1133-00 |
| 1049+05 | 7156-02 | 4278-02 | 2782-02 | 1910-02 | 1323-02 | 9471-03 | 7733-03 | 9603-03 | 1143-01 | 6884-01 | 8822-01 | 1291-00 |
| 1290+05 | 3161-01 | 1842-01 | 1173-01 | 8244-02 | 6139-02 | 4691-02 | 3938-02 | 4801-02 | 2820-01 | 9028-01 | 9932-01 | 1465-00 |
| 1587+05 | 1054-00 | 6067-01 | 3748-01 | 2574-01 | 1962-01 | 1579-01 | 1358-01 | 1553-01 | 4816-01 | 1060-00 | 1125-00 | 1657-00 |
| 1953+05 | 2677-00 | 1576-00 | 9606-01 | 6381-01 | 4776-01 | 3928-01 | 3416-01 | 3489-01 | 7121-01 | 1182-00 | 1279-00 | 1867-00 |
| 2402+05 | 5105-00 | 3253-00 | 2026-00 | 1321-00 | 9565-01 | 7768-01 | 6714-01 | 6105-01 | 9794-01 | 1312-00 | 1459-00 | 2095-00 |
| 2955+05 | 7353-00 | 5342-00 | 3557-00 | 2348-00 | 1660-00 | 1303-00 | 1103-00 | 9204-01 | 1279-00 | 1464-00 | 1665-00 | 2341-00 |
| 3635+05 | 8675-00 | 7164-00 | 5276-00 | 3647-00 | 2577-00 | 1952-00 | 1605-00 | 1270-00 | 1588-00 | 1644-00 | 1898-00 | 2606-00 |
| 4472+05 | 9290-00 | 8348-00 | 6793-00 | 5037-00 | 3652-00 | 2701-00 | 2156-00 | 1658-00 | 1877-00 | 1853-00 | 2159-00 | 2887-00 |
| 5501+05 | 9579-00 | 9011-00 | 7903-00 | 6316-00 | 4795-00 | 3526-00 | 2743-00 | 2083-00 | 2140-00 | 2094-00 | 2447-00 | 3187-00 |
| 6768+05 | 9730-00 | 9377-00 | 8631-00 | 7359-00 | 5900-00 | 4393-00 | 3360-00 | 2543-00 | 2401-00 | 2367-00 | 2761-00 | 3503-00 |
| 8326+05 | 9818-00 | 9590-00 | 9092-00 | 8140-00 | 6883-00 | 5267-00 | 4001-00 | 3037-00 | 2683-00 | 2672-00 | 3100-00 | 3835-00 |
| 1024+06 | 9874-00 | 9722-00 | 9386-00 | 8698-00 | 7697-00 | 6106-00 | 4659-00 | 3558-00 | 2996-00 | 3008-00 | 3460-00 | 4181-00 |
| 1260+06 | 9911-00 | 9807-00 | 9577-00 | 9088-00 | 8331-00 | 6876-00 | 5322-00 | 4102-00 | 3341-00 | 3372-00 | 3839-00 | 4539-00 |
| 1550+06 | 9937-00 | 9864-00 | 9705-00 | 9358-00 | 8806-00 | 7552-00 | 5975-00 | 4661-00 | 3718-00 | 3760-00 | 4233-00 | 4907-00 |
| 1907+06 | 9955-00 | 9903-00 | 9792-00 | 9546-00 | 9150-00 | 8120-00 | 6603-00 | 5230-00 | 4123-00 | 4167-00 | 4637-00 | 5281-00 |
| 2346+06 | 9967-00 | 9931-00 | 9852-00 | 9677-00 | 9397-00 | 8580-00 | 7189-00 | 5799-00 | 4552-00 | 4588-00 | 5046-00 | 5657-00 |
| 2885+06 | 9976-00 | 9950-00 | 9894-00 | 9769-00 | 9572-00 | 8941-00 | 7721-00 | 6358-00 | 5002-00 | 5019-00 | 5457-00 | 6031-00 |
| 3550+06 | 9983-00 | 9964-00 | 9924-00 | 9835-00 | 9695-00 | 9218-00 | 8187-00 | 6896-00 | 5465-00 | 5455-00 | 5864-00 | 6399-00 |
| 4367+06 | 9987-00 | 9974-00 | 9945-00 | 9881-00 | 9783-00 | 9426-00 | 8585-00 | 7402-00 | 5938-00 | 5890-00 | 6266-00 | 6759-00 |
| 5372+06 | 9991-00 | 9981-00 | 9960-00 | 9914-00 | 9845-00 | 9581-00 | 8914-00 | 7863-00 | 6411-00 | 6323-00 | 6658-00 | 7106-00 |
| 6608+06 | 9993-00 | 9986-00 | 9971-00 | 9938-00 | 9889-00 | 9694-00 | 9180-00 | 8272-00 | 6877-00 | 6749-00 | 7038-00 | 7440-00 |
| 8129+06 | 9995-00 | 9990-00 | 9979-00 | 9955-00 | 9920-00 | 9778-00 | 9389-00 | 8626-00 | 7327-00 | 7164-00 | 7404-00 | 7759-00 |
| 1000+07 | 9996-00 | 9993-00 | 9985-00 | 9967-00 | 9943-00 | 9838-00 | 9550-00 | 8922-00 | 7751-00 | 7562-00 | 7755-00 | 8060-00 |

Table 2 *Dissociation Fraction* $f=(n_D + n_{D+})/2n_0$

| T, K | Density, g/cm³ | | | | | | | | | | | |
|---|---|---|---|---|---|---|---|---|---|---|---|---|
| | 1000-03 | 4237-03 | 1795-02 | 7606-02 | 3223-01 | 1365+00 | 5785+00 | 2451+01 | 1038+02 | 2000+02 | 3000+02 | 4000+02 |
| 2000+04 | 1945-02 | 9453-03 | 4594-03 | 2232-03 | 1084-03 | 5269-04 | 2559-04 | 1243-04 | 6042-05 | 4353-05 | 1287-03 | 1000+01 |
| 2460+04 | 2449-02 | 1191-02 | 5787-03 | 2813-03 | 1368-03 | 6689-04 | 3332-04 | 1814-04 | 1679-04 | 5064-04 | 1432-02 | 1000+01 |
| 3027+04 | 1897-01 | 9262-02 | 4511-02 | 2195-02 | 1068-02 | 5222-03 | 2602-03 | 1416-03 | 1310-03 | 3914-03 | 1554-00 | 1000+01 |
| 3723+04 | 9726-01 | 4851-01 | 2387-01 | 1167-01 | 5696-02 | 2788-02 | 1389-02 | 7560-03 | 6961-03 | 2001-02 | 4776-00 | 1000+01 |
| 4580+04 | 3288-00 | 1769-00 | 9033-01 | 4497-01 | 2213-01 | 1088-01 | 5428-02 | 2951-02 | 2678-02 | 7120-02 | 9287-00 | 1000+01 |
| 5635+04 | 6851-00 | 4428-00 | 2496-00 | 1305-00 | 6578-01 | 3268-01 | 1637-01 | 8875-02 | 7797-02 | 5723-01 | 9951-00 | 1000+01 |
| 6931+04 | 9096-00 | 7439-00 | 5032-00 | 2918-00 | 1548-00 | 7878-01 | 3981-01 | 2149-01 | 1823-01 | 1874-00 | 9997-00 | 1000+01 |
| 8527+04 | 9765-00 | 9130-00 | 7505-00 | 5103-00 | 2967-00 | 1578-00 | 8112-01 | 4371-01 | 4045-01 | 3527-00 | 9999-00 | 1000+01 |
| 1049+05 | 9930-00 | 9715-00 | 8966-00 | 7181-00 | 4749-00 | 2712-00 | 1440-00 | 7862-01 | 9537-01 | 5781-00 | 1000+01 | 1000+01 |
| 1290+05 | 9976-00 | 9898-00 | 9589-00 | 8592-00 | 6530-00 | 4118-00 | 2307-00 | 1312-00 | 1768-00 | 8321-00 | 1000+01 | 1000+01 |
| 1587+05 | 9992-00 | 9962-00 | 9833-00 | 9345-00 | 7963-00 | 5645-00 | 3416-00 | 2053-00 | 2767-00 | 9673-00 | 1000+01 | 1000+01 |
| 1953+05 | 9998-00 | 9986-00 | 9933-00 | 9707-00 | 8915-00 | 7080-00 | 4699-00 | 2952-00 | 3968-00 | 9959-00 | 1000+01 | 1000+01 |
| 2402+05 | 1000+01 | 9996-00 | 9975-00 | 9875-00 | 9462-00 | 8221-00 | 5999-00 | 3898-00 | 5397-00 | 9991-00 | 1000+01 | 1000+01 |
| 2955+05 | 1000+01 | 9999-00 | 9992-00 | 9951-00 | 9747-00 | 8996-00 | 7151-00 | 4822-00 | 6973-00 | 9995-00 | 1000+01 | 1000+01 |
| 3635+05 | 1000+01 | 1000+01 | 9998-00 | 9982-00 | 9887-00 | 9461-00 | 8068-00 | 5698-00 | 8423-00 | 9996-00 | 1000+01 | 1000+01 |
| 4472+05 | 1000+01 | 1000+01 | 1000+01 | 9994-00 | 9952-00 | 9719-00 | 8741-00 | 6514-00 | 9401-00 | 9997-00 | 1000+01 | 1000+01 |
| 5501+05 | 1000+01 | 1000+01 | 1000+01 | 9998-00 | 9980-00 | 9857-00 | 9206-00 | 7258-00 | 9839-00 | 9998-00 | 1000+01 | 1000+01 |
| 6768+05 | 1000+01 | 1000+01 | 1000+01 | 1000+01 | 9992-00 | 9929-00 | 9513-00 | 7916-00 | 9967-00 | 9998-00 | 1000+01 | 1000+01 |
| 8326+05 | 1000+01 | 1000+01 | 1000+01 | 1000+01 | 9997-00 | 9965-00 | 9709-00 | 8475-00 | 9992-00 | 9998-00 | 1000+01 | 1000+01 |
| 1024+06 | 1000+01 | 1000+01 | 1000+01 | 1000+01 | 9999-00 | 9984-00 | 9830-00 | 8928-00 | 9997-00 | 9999-00 | 1000+01 | 1000+01 |
| 1260+06 | 1000+01 | 1000+01 | 1000+01 | 1000+01 | 1000+01 | 9992-00 | 9904-00 | 9273-00 | 9998-00 | 9999-00 | 1000+01 | 1000+01 |
| 1550+06 | 1000+01 | 1000+01 | 1000+01 | 1000+01 | 1000+01 | 9997-00 | 9947-00 | 9524-00 | 9998-00 | 9999-00 | 1000+01 | 1000+01 |
| 1907+06 | 1000+01 | 1000+01 | 1000+01 | 1000+01 | 1000+01 | 9999-00 | 9972-00 | 9697-00 | 9999-00 | 9999-00 | 1000+01 | 1000+01 |
| 2346+06 | 1000+01 | 1000+01 | 1000+01 | 1000+01 | 1000+01 | 9999-00 | 9986-00 | 9813-00 | 9999-00 | 9999-00 | 1000+01 | 1000+01 |
| 2885+06 | 1000+01 | 1000+01 | 1000+01 | 1000+01 | 1000+01 | 1000+01 | 9993-00 | 9887-00 | 9999-00 | 1000+01 | 1000+01 | 1000+01 |
| 3550+06 | 1000+01 | 1000+01 | 1000+01 | 1000+01 | 1000+01 | 1000+01 | 9997-00 | 9935-00 | 1000+01 | 1000+01 | 1000+01 | 1000+01 |
| 4367+06 | 1000+01 | 1000+01 | 1000+01 | 1000+01 | 1000+01 | 1000+01 | 9998-00 | 9963-00 | 1000+01 | 1000+01 | 1000+01 | 1000+01 |
| 5372+06 | 1000+01 | 1000+01 | 1000+01 | 1000+01 | 1000+01 | 1000+01 | 9999-00 | 9980-00 | 1000+01 | 1000+01 | 1000+01 | 1000+01 |
| 6608+06 | 1000+01 | 1000+01 | 1000+01 | 1000+01 | 1000+01 | 1000+01 | 1000+01 | 9990-00 | 1000+01 | 1000+01 | 1000+01 | 1000+01 |
| 8129+06 | 1000+01 | 1000+01 | 1000+01 | 1000+01 | 1000+01 | 1000+01 | 1000+01 | 9995-00 | 1000+01 | 1000+01 | 1000+01 | 1000+01 |
| 1000+07 | 1000+01 | 1000+01 | 1000+01 | 1000+01 | 1000+01 | 1000+01 | 1000+01 | 9998-00 | 1000+01 | 1000+01 | 1000+01 | 1000+01 |

Table 3 *Pressure in Pa*

| T, K | Density, g/cm³ | | | | | | | | | | | |
|---|---|---|---|---|---|---|---|---|---|---|---|---|
| | 1000-03 | 4237-03 | 1795-02 | 7606-02 | 3223-01 | 1365+00 | 5785+00 | 2451+01 | 1038+02 | 2000+02 | 3000+02 | 4000+02 |
| 2000+04 | 4136+06 | 1751+07 | 7414+07 | 3142+08 | 1334+09 | 5695+09 | 2499+10 | 1231+11 | 1046+12 | 5411+12 | 2927+13 | 4400+14 |
| 2460+04 | 5090+06 | 2154+07 | 9122+07 | 3866+08 | 1641+09 | 7006+09 | 3075+10 | 1515+11 | 1286+12 | 6627+12 | 3474+13 | 5120+14 |
| 3027+04 | 6365+06 | 2671+07 | 1127+08 | 4765+08 | 2021+09 | 8623+09 | 3783+10 | 1864+11 | 1582+12 | 8083+12 | 4293+13 | 5913+14 |
| 3723+04 | 8432+06 | 3414+07 | 1413+08 | 5917+08 | 2497+09 | 1063+10 | 4659+10 | 2294+11 | 1944+12 | 9775+12 | 6680+13 | 6770+14 |
| 4580+04 | 1256+07 | 4714+07 | 1850+08 | 7518+08 | 3122+09 | 1318+10 | 5755+10 | 2828+11 | 2387+12 | 1161+13 | 1139+14 | 7681+14 |
| 5635+04 | 1960+07 | 7110+07 | 2609+08 | 1001+09 | 4005+09 | 1657+10 | 7157+10 | 3500+11 | 2930+12 | 1333+13 | 1432+14 | 8631+14 |
| 6931+04 | 2732+07 | 1057+08 | 3862+08 | 1407+09 | 5342+09 | 2131+10 | 9015+10 | 4362+11 | 3582+12 | 1585+13 | 1712+14 | 9606+14 |
| 8527+04 | 3482+07 | 1427+08 | 5535+08 | 2026+09 | 7391+09 | 2820+10 | 1156+11 | 5496+11 | 4327+12 | 2031+13 | 2031+14 | 1059+15 |
| 1049+05 | 4339+07 | 1814+08 | 7386+08 | 2838+09 | 1037+10 | 3825+10 | 1513+11 | 7019+11 | 5086+12 | 2948+13 | 2397+14 | 1157+15 |
| 1290+05 | 5457+07 | 2275+08 | 9431+08 | 3788+09 | 1434+10 | 5255+10 | 2019+11 | 9095+11 | 5793+12 | 4768+13 | 2812+14 | 1252+15 |
| 1587+05 | 7139+07 | 2899+08 | 1194+09 | 4889+09 | 1925+10 | 7207+10 | 2733+11 | 1194+12 | 6507+12 | 6944+13 | 3278+14 | 1345+15 |
| 1953+05 | 9980+07 | 3853+08 | 1540+09 | 6264+09 | 2522+10 | 9754+10 | 3729+11 | 1582+12 | 7588+12 | 8782+13 | 3791+14 | 1433+15 |
| 2402+05 | 1457+08 | 5383+08 | 2058+09 | 8144+09 | 3274+10 | 1298+11 | 5085+11 | 2102+12 | 1004+13 | 1059+14 | 4349+14 | 1517+15 |
| 2955+05 | 2064+08 | 7662+08 | 2843+09 | 1085+10 | 4282+10 | 1704+11 | 6885+11 | 2783+12 | 1527+13 | 1264+14 | 4947+14 | 1595+15 |
| 3635+05 | 2742+08 | 1058+09 | 3950+09 | 1473+10 | 5680+10 | 2226+11 | 9224+11 | 3663+12 | 2392+13 | 1500+14 | 5578+14 | 1669+15 |
| 4472+05 | 3495+08 | 1398+09 | 5369+09 | 2004+10 | 7619+10 | 2910+11 | 1222+12 | 4794+12 | 3538+13 | 1773+14 | 6237+14 | 1738+15 |
| 5501+05 | 4376+08 | 1788+09 | 7076+09 | 2691+10 | 1024+11 | 3815+11 | 1604+12 | 6245+12 | 4789+13 | 2084+14 | 6921+14 | 1803+15 |
| 6768+05 | 5438+08 | 2249+09 | 9097+09 | 3542+10 | 1365+11 | 5011+11 | 2092+12 | 8110+12 | 6069+13 | 2438+14 | 7627+14 | 1865+15 |
| 8326+05 | 6734+08 | 2805+09 | 1151+10 | 4577+10 | 1797+11 | 6575+11 | 2712+12 | 1051+13 | 7462+13 | 2837+14 | 8359+14 | 1926+15 |
| 1024+06 | 8322+08 | 3483+09 | 1442+10 | 5830+10 | 2331+11 | 8596+11 | 3504+12 | 1359+13 | 9093+13 | 3286+14 | 9125+14 | 1988+15 |
| 1260+06 | 1027+09 | 4313+09 | 1797+10 | 7352+10 | 2984+11 | 1117+12 | 4510+12 | 1752+13 | 1106+14 | 3790+14 | 9939+14 | 2054+15 |
| 1550+06 | 1267+09 | 5331+09 | 2230+10 | 9207+10 | 3779+11 | 1442+12 | 5787+12 | 2253+13 | 1346+14 | 4361+14 | 1082+15 | 2130+15 |
| 1907+06 | 1562+09 | 6581+09 | 2762+10 | 1147+11 | 4749+11 | 1846+12 | 7399+12 | 2887+13 | 1636+14 | 5016+14 | 1178+15 | 2218+15 |
| 2346+06 | 1924+09 | 8119+09 | 3414+10 | 1425+11 | 5933+11 | 2346+12 | 9425+12 | 3688+13 | 1988+14 | 5783+14 | 1286+15 | 2326+15 |
| 2885+06 | 2370+09 | 1001+10 | 4216+10 | 1766+11 | 7383+11 | 2963+12 | 1196+13 | 4696+13 | 2417+14 | 6693+14 | 1411+15 | 2456+15 |
| 3550+06 | 2919+09 | 1233+10 | 5202+10 | 2185+11 | 9161+11 | 3721+12 | 1511+13 | 5961+13 | 2945+14 | 7781+14 | 1562+15 | 2615+15 |
| 4367+06 | 3593+09 | 1519+10 | 6414+10 | 2699+11 | 1134+12 | 4652+12 | 1901+13 | 7547+13 | 3596+14 | 9076+14 | 1747+15 | 2820+15 |
| 5372+06 | 4423+09 | 1871+10 | 7904+10 | 3331+11 | 1402+12 | 5795+12 | 2384+13 | 9527+13 | 4403+14 | 1064+15 | 1977+15 | 3089+15 |
| 6608+06 | 5444+09 | 2304+10 | 9737+10 | 4109+11 | 1732+12 | 7198+12 | 2978+13 | 1199+14 | 5403+14 | 1255+15 | 2254+15 | 3438+15 |
| 8129+06 | 6700+09 | 2836+10 | 1199+11 | 5064+11 | 2137+12 | 8921+12 | 3710+13 | 1505+14 | 6646+14 | 1490+15 | 2593+15 | 3873+15 |
| 1000+07 | 8245+09 | 3490+10 | 1477+11 | 6240+11 | 2635+12 | 1104+13 | 4609+13 | 1883+14 | 8190+14 | 1780+15 | 3014+15 | 4409+15 |

Table 4 *Specific internal energy in J/kg*

| T, K | Density, g/cm³ | | | | | | | | | | | |
|---|---|---|---|---|---|---|---|---|---|---|---|---|
| | 1000-03 | 4237-03 | 1795-02 | 7606-02 | 3223-01 | 1365+00 | 5785+00 | 2451+01 | 1038+02 | 2000+02 | 3000+02 | 4000+02 |
| 2000+04 | -9843+08 | -9854+08 | -9860+08 | -9862+08 | -9863+08 | -9864+08 | -9864+08 | -9865+08 | -9865+08 | -9872+08 | -9993+08 | 2530+09 |
| 2460+04 | -9588+08 | -9602+08 | -9609+08 | -9612+08 | -9614+08 | -9615+08 | -9615+08 | -9615+08 | -9616+08 | -9632+08 | -9912+08 | 3062+09 |
| 3027+04 | -9090+08 | -9198+08 | -9251+08 | -9277+08 | -9290+08 | -9296+08 | -9299+08 | -9300+08 | -9302+08 | -9340+08 | -9530+08 | 3718+09 |
| 3723+04 | -7813+08 | -8359+08 | -8636+08 | -8772+08 | -8839+08 | -8872+08 | -8888+08 | -8895+08 | -8900+08 | -8993+08 | -3677+08 | 4518+09 |
| 4580+04 | -4704+08 | -6411+08 | -7383+08 | -7892+08 | -8149+08 | -8275+08 | -8336+08 | -8364+08 | -8382+08 | -8616+08 | 8397+08 | 5474+09 |
| 5635+04 | -5571+06 | -2785+08 | -4959+08 | -6300+08 | -7029+08 | -7402+08 | -7585+08 | -7670+08 | -7722+08 | -8081+08 | 1234+09 | 6600+09 |
| 6931+04 | 3278+08 | 1399+08 | -1326+08 | -3719+08 | -5269+08 | -6130+08 | -6571+08 | -6779+08 | -6922+08 | -7257+08 | 1524+09 | 7899+09 |
| 8527+04 | 5090+08 | 4330+08 | 2452+08 | -3084+07 | -2760+08 | -4354+08 | -5234+08 | -5664+08 | -5971+08 | -5563+08 | 1876+09 | 9373+09 |
| 1049+05 | 6907+08 | 6449+08 | 5464+08 | 3310+08 | 4195+07 | -1993+08 | -3497+08 | -4269+08 | -4767+08 | -1555+08 | 2323+09 | 1102+10 |
| 1290+05 | 1021+09 | 9148+08 | 8274+08 | 6789+08 | 4112+08 | 1065+08 | -1207+08 | -2455+08 | -3497+08 | 6027+08 | 2888+09 | 1282+10 |
| 1587+05 | 1762+09 | 1419+09 | 1223+09 | 1065+09 | 8318+08 | 4984+08 | 1883+08 | -3796+06 | -2174+08 | 1359+09 | 3596+09 | 1477+10 |
| 1953+05 | 3284+09 | 2412+09 | 1908+09 | 1603+09 | 1337+09 | 9917+08 | 5996+08 | 3055+08 | -4238+07 | 1901+09 | 4476+09 | 1686+10 |
| 2402+05 | 5624+09 | 4104+09 | 3065+09 | 2424+09 | 1988+09 | 1596+09 | 1121+09 | 6859+08 | 2745+08 | 2433+09 | 5556+09 | 1907+10 |
| 2955+05 | 8028+09 | 6332+09 | 4760+09 | 3629+09 | 2858+09 | 2327+09 | 1757+09 | 1153+09 | 8621+08 | 3082+09 | 6862+09 | 2140+10 |
| 3635+05 | 9862+09 | 8550+09 | 6819+09 | 5228+09 | 4008+09 | 3220+09 | 2522+09 | 1735+09 | 1766+09 | 3899+09 | 8421+09 | 2384+10 |
| 4472+05 | 1133+10 | 1045+10 | 8946+09 | 7136+09 | 5476+09 | 4326+09 | 3447+09 | 2470+09 | 2904+09 | 4929+09 | 1026+10 | 2639+10 |
| 5501+05 | 1278+10 | 1218+10 | 1100+10 | 9238+09 | 7276+09 | 5707+09 | 4581+09 | 3406+09 | 4136+09 | 6220+09 | 1239+10 | 2907+10 |
| 6768+05 | 1442+10 | 1398+10 | 1306+10 | 1149+10 | 9410+09 | 7428+09 | 5989+09 | 4605+09 | 5451+09 | 7826+09 | 1485+10 | 3189+10 |
| 8326+05 | 1639+10 | 1603+10 | 1529+10 | 1392+10 | 1189+10 | 9560+09 | 7750+09 | 6141+09 | 6978+09 | 9806+09 | 1767+10 | 3488+10 |
| 1024+06 | 1879+10 | 1847+10 | 1785+10 | 1666+10 | 1476+10 | 1218+10 | 9964+09 | 8103+09 | 8870+09 | 1222+10 | 2090+10 | 3811+10 |
| 1260+06 | 2173+10 | 2143+10 | 2088+10 | 1984+10 | 1810+10 | 1538+10 | 1275+10 | 1060+10 | 1126+10 | 1515+10 | 2460+10 | 4165+10 |
| 1550+06 | 2533+10 | 2506+10 | 2455+10 | 2362+10 | 2204+10 | 1928+10 | 1624+10 | 1376+10 | 1429+10 | 1868+10 | 2885+10 | 4562+10 |
| 1907+06 | 2976+10 | 2950+10 | 2903+10 | 2817+10 | 2673+10 | 2400+10 | 2061+10 | 1776+10 | 1810+10 | 2297+10 | 3370+10 | 5017+10 |
| 2346+06 | 3520+10 | 3495+10 | 3451+10 | 3371+10 | 3238+10 | 2974+10 | 2606+10 | 2279+10 | 2284+10 | 2824+10 | 3929+10 | 5544+10 |
| 2885+06 | 4189+10 | 4165+10 | 4123+10 | 4047+10 | 3924+10 | 3670+10 | 3282+10 | 2911+10 | 2879+10 | 3473+10 | 4585+10 | 6153+10 |
| 3550+06 | 5012+10 | 4988+10 | 4948+10 | 4876+10 | 4760+10 | 4517+10 | 4118+10 | 3703+10 | 3623+10 | 4271+10 | 5381+10 | 6870+10 |
| 4367+06 | 6024+10 | 6001+10 | 5962+10 | 5893+10 | 5782+10 | 5551+10 | 5149+10 | 4691+10 | 4556+10 | 5238+10 | 6362+10 | 7751+10 |
| 5372+06 | 7269+10 | 7247+10 | 7208+10 | 7142+10 | 7036+10 | 6816+10 | 6416+10 | 5922+10 | 5723+10 | 6421+10 | 7570+10 | 8872+10 |
| 6608+06 | 8801+10 | 8778+10 | 8741+10 | 8677+10 | 8574+10 | 8364+10 | 7972+10 | 7447+10 | 7181+10 | 7880+10 | 9025+10 | 1029+11 |
| 8129+06 | 1068+11 | 1066+11 | 1063+11 | 1056+11 | 1046+11 | 1026+11 | 9883+10 | 9333+10 | 8998+10 | 9686+10 | 1080+11 | 1202+11 |
| 1000+07 | 1300+11 | 1298+11 | 1294+11 | 1288+11 | 1279+11 | 1259+11 | 1223+11 | 1166+11 | 1126+11 | 1192+11 | 1300+11 | 1414+11 |

Table 5 *Specific heat at constant pressure, cp, in J/kg-K*

| T, K | Density, g/cm³ | | | | | | | | | | | |
|---|---|---|---|---|---|---|---|---|---|---|---|---|
| | 1000-03 | 4237-03 | 1795-02 | 7606-02 | 3223-01 | 1365+00 | 5785+00 | 2451+01 | 1038+02 | 2000+02 | 3000+02 | 4000+02 |
| 2000+04 | 9147+04 | 8268+04 | 7841+04 | 7633+04 | 7532+04 | 7483+04 | 7461+04 | 7476+04 | 7918+04 | 9255+04 | 9682+04 | 1474+06 |
| 2460+04 | 8902+04 | 8197+04 | 7855+04 | 7688+04 | 7607+04 | 7568+04 | 7551+04 | 7568+04 | 8003+04 | 9106+04 | 6044+04 | 1451+06 |
| 3027+04 | 1481+05 | 1116+05 | 9379+04 | 8508+04 | 8085+04 | 7879+04 | 7782+04 | 7763+04 | 8168+04 | 8806+04 | 4820+05 | 1412+06 |
| 3723+04 | 3233+05 | 2019+05 | 1396+05 | 1085+05 | 9330+04 | 8589+04 | 8234+04 | 8098+04 | 8430+04 | 8099+04 | 1867+06 | 1355+06 |
| 4580+04 | 5861+05 | 3770+05 | 2358+05 | 1583+05 | 1187+05 | 9913+04 | 8967+04 | 8559+04 | 8710+04 | 6357+04 | 1001+06 | 1281+06 |
| 5635+04 | 5215+05 | 5061+05 | 3601+05 | 2336+05 | 1583+05 | 1189+05 | 9934+04 | 9053+04 | 8778+04 | 8008+04 | 3788+05 | 1191+06 |
| 6931+04 | 2336+05 | 3695+05 | 3969+05 | 3028+05 | 2046+05 | 1429+05 | 1103+05 | 9492+04 | 8422+04 | 1083+05 | 3344+05 | 1090+06 |
| 8527+04 | 1396+05 | 1964+05 | 2903+05 | 3097+05 | 2403+05 | 1676+05 | 1222+05 | 9942+04 | 8137+04 | 1989+05 | 3317+05 | 9814+05 |
| 1049+05 | 1516+05 | 1504+05 | 1905+05 | 2521+05 | 2489+05 | 1900+05 | 1368+05 | 1069+05 | 7332+04 | 3760+05 | 3302+05 | 8705+05 |
| 1290+05 | 2386+05 | 1838+05 | 1712+05 | 2002+05 | 2324+05 | 2074+05 | 1562+05 | 1195+05 | 5822+04 | 4665+05 | 3268+05 | 7619+05 |
| 1587+05 | 4243+05 | 2827+05 | 2129+05 | 1939+05 | 2124+05 | 2163+05 | 1778+05 | 1330+05 | 5433+04 | 2909+05 | 3207+05 | 6592+05 |
| 1953+05 | 6643+05 | 4381+05 | 2983+05 | 2272+05 | 2077+05 | 2148+05 | 1942+05 | 1423+05 | 7642+04 | 1913+05 | 3118+05 | 5650+05 |
| 2402+05 | 7300+05 | 5740+05 | 4002+05 | 2822+05 | 2215+05 | 2082+05 | 2010+05 | 1490+05 | 1341+05 | 1725+05 | 2999+05 | 4809+05 |
| 2955+05 | 5259+05 | 5643+05 | 4631+05 | 3358+05 | 2474+05 | 2047+05 | 2003+05 | 1556+05 | 1965+05 | 1712+05 | 2852+05 | 4074+05 |
| 3635+05 | 3340+05 | 4308+05 | 4432+05 | 3632+05 | 2756+05 | 2078+05 | 1971+05 | 1632+05 | 2237+05 | 1710+05 | 2683+05 | 3446+05 |
| 4472+05 | 2528+05 | 3140+05 | 3695+05 | 3541+05 | 2961+05 | 2159+05 | 1946+05 | 1718+05 | 2057+05 | 1704+05 | 2501+05 | 2918+05 |
| 5501+05 | 2253+05 | 2549+05 | 3005+05 | 3215+05 | 3024+05 | 2260+05 | 1944+05 | 1810+05 | 1700+05 | 1690+05 | 2314+05 | 2482+05 |
| 6768+05 | 2155+05 | 2296+05 | 2574+05 | 2856+05 | 2944+05 | 2352+05 | 1966+05 | 1898+05 | 1488+05 | 1668+05 | 2132+05 | 2127+05 |
| 8326+05 | 2116+05 | 2188+05 | 2345+05 | 2577+05 | 2782+05 | 2416+05 | 2005+05 | 1972+05 | 1439+05 | 1634+05 | 1966+05 | 1847+05 |
| 1024+06 | 2097+05 | 2137+05 | 2228+05 | 2392+05 | 2603+05 | 2444+05 | 2052+05 | 2025+05 | 1464+05 | 1592+05 | 1819+05 | 1631+05 |
| 1260+06 | 2087+05 | 2111+05 | 2165+05 | 2276+05 | 2449+05 | 2437+05 | 2100+05 | 2060+05 | 1504+05 | 1547+05 | 1693+05 | 1475+05 |
| 1550+06 | 2080+05 | 2096+05 | 2130+05 | 2205+05 | 2332+05 | 2403+05 | 2143+05 | 2085+05 | 1544+05 | 1516+05 | 1580+05 | 1370+05 |
| 1907+06 | 2076+05 | 2087+05 | 2110+05 | 2160+05 | 2249+05 | 2355+05 | 2176+05 | 2106+05 | 1566+05 | 1510+05 | 1478+05 | 1299+05 |
| 2346+06 | 2073+05 | 2081+05 | 2096+05 | 2131+05 | 2192+05 | 2302+05 | 2196+05 | 2126+05 | 1594+05 | 1525+05 | 1405+05 | 1239+05 |
| 2885+06 | 2071+05 | 2077+05 | 2088+05 | 2112+05 | 2154+05 | 2253+05 | 2203+05 | 2145+05 | 1637+05 | 1549+05 | 1385+05 | 1189+05 |
| 3550+06 | 2069+05 | 2074+05 | 2082+05 | 2099+05 | 2128+05 | 2211+05 | 2201+05 | 2160+05 | 1687+05 | 1558+05 | 1418+05 | 1184+05 |
| 4367+06 | 2068+05 | 2072+05 | 2077+05 | 2090+05 | 2111+05 | 2177+05 | 2190+05 | 2170+05 | 1740+05 | 1555+05 | 1465+05 | 1251+05 |
| 5372+06 | 2067+05 | 2070+05 | 2074+05 | 2084+05 | 2099+05 | 2150+05 | 2175+05 | 2175+05 | 1794+05 | 1588+05 | 1492+05 | 1345+05 |
| 6608+06 | 2067+05 | 2069+05 | 2072+05 | 2079+05 | 2090+05 | 2129+05 | 2158+05 | 2174+05 | 1848+05 | 1634+05 | 1495+05 | 1424+05 |
| 8129+06 | 2066+05 | 2068+05 | 2070+05 | 2076+05 | 2084+05 | 2113+05 | 2142+05 | 2168+05 | 1898+05 | 1686+05 | 1544+05 | 1438+05 |
| 1000+07 | 2066+05 | 2067+05 | 2069+05 | 2073+05 | 2079+05 | 2101+05 | 2127+05 | 2158+05 | 1943+05 | 1740+05 | 1606+05 | 1500+05 |

Table 6 *Ratio of specific heats, cp/cv*

| T, K | Density, g/cm³ | | | | | | | | | | | |
|---|---|---|---|---|---|---|---|---|---|---|---|---|
| | 1000-03 | 4237-03 | 1795-02 | 7606-02 | 3223-01 | 1365+00 | 5785+00 | 2451+01 | 1038+02 | 2000+02 | 3000+02 | 4000+02 |
| 2000+04 | 1315+01 | 1346+01 | 1364+01 | 1374+01 | 1379+01 | 1382+01 | 1383+01 | 1389+01 | 1474+01 | 1771+01 | 3550+01 | 1284+01 |
| 2460+04 | 1326+01 | 1350+01 | 1363+01 | 1370+01 | 1374+01 | 1376+01 | 1377+01 | 1382+01 | 1465+01 | 1755+01 | 8572+01 | 1251+01 |
| 3027+04 | 1243+01 | 1282+01 | 1316+01 | 1339+01 | 1353+01 | 1360+01 | 1364+01 | 1370+01 | 1450+01 | 1726+01 | 1469+01 | 1221+01 |
| 3723+04 | 1237+01 | 1242+01 | 1265+01 | 1295+01 | 1319+01 | 1336+01 | 1345+01 | 1354+01 | 1428+01 | 1683+01 | 1282+01 | 1194+01 |
| 4580+04 | 1325+01 | 1291+01 | 1274+01 | 1279+01 | 1296+01 | 1315+01 | 1329+01 | 1340+01 | 1408+01 | 1635+01 | 1356+01 | 1169+01 |
| 5635+04 | 1382+01 | 1388+01 | 1354+01 | 1323+01 | 1313+01 | 1319+01 | 1329+01 | 1341+01 | 1400+01 | 1283+01 | 1538+01 | 1147+01 |
| 6931+04 | 1439+01 | 1436+01 | 1448+01 | 1418+01 | 1380+01 | 1361+01 | 1359+01 | 1366+01 | 1399+01 | 1417+01 | 1532+01 | 1128+01 |
| 8527+04 | 1533+01 | 1493+01 | 1496+01 | 1510+01 | 1480+01 | 1442+01 | 1420+01 | 1416+01 | 1339+01 | 1383+01 | 1482+01 | 1111+01 |
| 1049+05 | 1452+01 | 1499+01 | 1514+01 | 1547+01 | 1565+01 | 1537+01 | 1502+01 | 1476+01 | 1253+01 | 1407+01 | 1431+01 | 1096+01 |
| 1290+05 | 1330+01 | 1395+01 | 1458+01 | 1518+01 | 1582+01 | 1603+01 | 1578+01 | 1528+01 | 1223+01 | 1467+01 | 1382+01 | 1083+01 |
| 1587+05 | 1300+01 | 1325+01 | 1375+01 | 1445+01 | 1536+01 | 1614+01 | 1636+01 | 1589+01 | 1244+01 | 1527+01 | 1336+01 | 1072+01 |
| 1953+05 | 1355+01 | 1338+01 | 1351+01 | 1396+01 | 1483+01 | 1588+01 | 1686+01 | 1676+01 | 1377+01 | 1525+01 | 1294+01 | 1062+01 |
| 2402+05 | 1434+01 | 1409+01 | 1392+01 | 1404+01 | 1472+01 | 1562+01 | 1732+01 | 1765+01 | 1538+01 | 1484+01 | 1256+01 | 1053+01 |
| 2955+05 | 1500+01 | 1494+01 | 1475+01 | 1461+01 | 1513+01 | 1558+01 | 1761+01 | 1835+01 | 1609+01 | 1443+01 | 1222+01 | 1046+01 |
| 3635+05 | 1585+01 | 1568+01 | 1563+01 | 1543+01 | 1591+01 | 1580+01 | 1771+01 | 1887+01 | 1615+01 | 1406+01 | 1194+01 | 1039+01 |
| 4472+05 | 1656+01 | 1637+01 | 1633+01 | 1624+01 | 1680+01 | 1622+01 | 1767+01 | 1925+01 | 1585+01 | 1370+01 | 1170+01 | 1034+01 |
| 5501+05 | 1682+01 | 1681+01 | 1681+01 | 1684+01 | 1753+01 | 1673+01 | 1760+01 | 1952+01 | 1536+01 | 1339+01 | 1150+01 | 1030+01 |
| 6768+05 | 1686+01 | 1695+01 | 1704+01 | 1719+01 | 1792+01 | 1723+01 | 1756+01 | 1966+01 | 1494+01 | 1312+01 | 1135+01 | 1028+01 |
| 8326+05 | 1684+01 | 1695+01 | 1709+01 | 1731+01 | 1800+01 | 1764+01 | 1755+01 | 1963+01 | 1473+01 | 1289+01 | 1124+01 | 1027+01 |
| 1024+06 | 1681+01 | 1690+01 | 1705+01 | 1730+01 | 1788+01 | 1791+01 | 1758+01 | 1945+01 | 1465+01 | 1271+01 | 1118+01 | 1028+01 |
| 1260+06 | 1678+01 | 1686+01 | 1698+01 | 1722+01 | 1768+01 | 1803+01 | 1761+01 | 1917+01 | 1461+01 | 1259+01 | 1116+01 | 1031+01 |
| 1550+06 | 1676+01 | 1682+01 | 1692+01 | 1712+01 | 1748+01 | 1801+01 | 1763+01 | 1887+01 | 1461+01 | 1254+01 | 1117+01 | 1038+01 |
| 1907+06 | 1674+01 | 1679+01 | 1686+01 | 1703+01 | 1730+01 | 1790+01 | 1763+01 | 1859+01 | 1459+01 | 1258+01 | 1123+01 | 1049+01 |
| 2346+06 | 1672+01 | 1676+01 | 1682+01 | 1696+01 | 1716+01 | 1774+01 | 1760+01 | 1835+01 | 1463+01 | 1269+01 | 1134+01 | 1063+01 |
| 2885+06 | 1671+01 | 1674+01 | 1679+01 | 1689+01 | 1705+01 | 1757+01 | 1754+01 | 1815+01 | 1475+01 | 1286+01 | 1153+01 | 1081+01 |
| 3550+06 | 1670+01 | 1673+01 | 1676+01 | 1685+01 | 1696+01 | 1740+01 | 1746+01 | 1798+01 | 1491+01 | 1304+01 | 1182+01 | 1106+01 |
| 4367+06 | 1670+01 | 1671+01 | 1674+01 | 1681+01 | 1690+01 | 1725+01 | 1736+01 | 1783+01 | 1511+01 | 1321+01 | 1217+01 | 1144+01 |
| 5372+06 | 1669+01 | 1671+01 | 1673+01 | 1678+01 | 1685+01 | 1713+01 | 1726+01 | 1769+01 | 1533+01 | 1348+01 | 1251+01 | 1190+01 |
| 6608+06 | 1669+01 | 1670+01 | 1672+01 | 1676+01 | 1681+01 | 1703+01 | 1716+01 | 1757+01 | 1556+01 | 1380+01 | 1280+01 | 1238+01 |
| 8129+06 | 1668+01 | 1669+01 | 1671+01 | 1674+01 | 1678+01 | 1695+01 | 1707+01 | 1745+01 | 1580+01 | 1415+01 | 1321+01 | 1273+01 |
| 1000+07 | 1668+01 | 1669+01 | 1670+01 | 1672+01 | 1676+01 | 1689+01 | 1699+01 | 1733+01 | 1601+01 | 1451+01 | 1364+01 | 1319+01 |

Table 7 *Sound speed in m/s*

| T, K | Density, g/cm³ | | | | | | | | | | | |
|---|---|---|---|---|---|---|---|---|---|---|---|---|
| | 1000-03 | 4237-03 | 1795-02 | 7606-02 | 3223-01 | 1365+00 | 5785+00 | 2451+01 | 1038+02 | 2000+02 | 3000+02 | 4000+02 |
| 2000+04 | 2331+04 | 2358+04 | 2374+04 | 2383+04 | 2392+04 | 2414+04 | 2500+04 | 2894+04 | 5416+04 | 1249+05 | 4324+05 | 1157+06 |
| 2460+04 | 2596+04 | 2619+04 | 2632+04 | 2640+04 | 2648+04 | 2672+04 | 2766+04 | 3203+04 | 5986+04 | 1373+05 | 7166+05 | 1208+06 |
| 3027+04 | 2800+04 | 2837+04 | 2871+04 | 2896+04 | 2915+04 | 2946+04 | 3054+04 | 3537+04 | 6598+04 | 1495+05 | 3916+05 | 1253+06 |
| 3723+04 | 3161+04 | 3127+04 | 3137+04 | 3166+04 | 3197+04 | 3240+04 | 3365+04 | 3900+04 | 7250+04 | 1604+05 | 5808+05 | 1291+06 |
| 4580+04 | 3872+04 | 3660+04 | 3552+04 | 3519+04 | 3529+04 | 3574+04 | 3714+04 | 4308+04 | 7956+04 | 1681+05 | 6342+05 | 1322+06 |
| 5635+04 | 4945+04 | 4553+04 | 4243+04 | 4060+04 | 3984+04 | 3992+04 | 4132+04 | 4789+04 | 8732+04 | 1565+05 | 6229+05 | 1345+06 |
| 6931+04 | 6145+04 | 5719+04 | 5263+04 | 4878+04 | 4640+04 | 4554+04 | 4665+04 | 5384+04 | 9511+04 | 2006+05 | 6592+05 | 1359+06 |
| 8527+04 | 7263+04 | 6954+04 | 6495+04 | 5982+04 | 5553+04 | 5316+04 | 5358+04 | 6130+04 | 9890+04 | 2535+05 | 6981+05 | 1367+06 |
| 1049+05 | 7913+04 | 7950+04 | 7715+04 | 7245+04 | 6704+04 | 6303+04 | 6238+04 | 7032+04 | 9715+04 | 3534+05 | 7362+05 | 1367+06 |
| 1290+05 | 8465+04 | 8606+04 | 8657+04 | 8446+04 | 7973+04 | 7480+04 | 7305+04 | 8071+04 | 9542+04 | 4649+05 | 7721+05 | 1362+06 |
| 1587+05 | 9449+04 | 9415+04 | 9477+04 | 9477+04 | 9234+04 | 8796+04 | 8578+04 | 9310+04 | 1010+05 | 4730+05 | 8054+05 | 1351+06 |
| 1953+05 | 1116+05 | 1075+05 | 1059+05 | 1057+05 | 1051+05 | 1025+05 | 1013+05 | 1081+05 | 1315+05 | 4635+05 | 8354+05 | 1337+06 |
| 2402+05 | 1370+05 | 1280+05 | 1227+05 | 1204+05 | 1197+05 | 1187+05 | 1202+05 | 1249+05 | 1974+05 | 4819+05 | 8619+05 | 1319+06 |
| 2955+05 | 1687+05 | 1562+05 | 1464+05 | 1405+05 | 1379+05 | 1373+05 | 1420+05 | 1426+05 | 2750+05 | 5118+05 | 8846+05 | 1298+06 |
| 3635+05 | 2034+05 | 1899+05 | 1770+05 | 1668+05 | 1609+05 | 1586+05 | 1663+05 | 1613+05 | 3447+05 | 5430+05 | 9036+05 | 1276+06 |
| 4472+05 | 2372+05 | 2260+05 | 2123+05 | 1990+05 | 1892+05 | 1835+05 | 1928+05 | 1817+05 | 3898+05 | 5738+05 | 9191+05 | 1253+06 |
| 5501+05 | 2690+05 | 2616+05 | 2499+05 | 2355+05 | 2225+05 | 2123+05 | 2217+05 | 2048+05 | 4030+05 | 6042+05 | 9316+05 | 1229+06 |
| 6768+05 | 3010+05 | 2965+05 | 2879+05 | 2747+05 | 2600+05 | 2453+05 | 2534+05 | 2318+05 | 4101+05 | 6343+05 | 9415+05 | 1205+06 |
| 8326+05 | 3353+05 | 3323+05 | 3264+05 | 3154+05 | 3009+05 | 2827+05 | 2885+05 | 2637+05 | 4298+05 | 6642+05 | 9496+05 | 1183+06 |
| 1024+06 | 3728+05 | 3706+05 | 3664+05 | 3579+05 | 3447+05 | 3246+05 | 3273+05 | 3007+05 | 4609+05 | 6940+05 | 9570+05 | 1162+06 |
| 1260+06 | 4142+05 | 4125+05 | 4093+05 | 4027+05 | 3914+05 | 3710+05 | 3705+05 | 3430+05 | 4981+05 | 7241+05 | 9647+05 | 1143+06 |
| 1550+06 | 4599+05 | 4585+05 | 4560+05 | 4508+05 | 4414+05 | 4220+05 | 4185+05 | 3905+05 | 5393+05 | 7548+05 | 9743+05 | 1128+06 |
| 1907+06 | 5106+05 | 5094+05 | 5074+05 | 5032+05 | 4954+05 | 4777+05 | 4718+05 | 4434+05 | 5846+05 | 7871+05 | 9865+05 | 1118+06 |
| 2346+06 | 5667+05 | 5656+05 | 5639+05 | 5605+05 | 5541+05 | 5384+05 | 5310+05 | 5022+05 | 6337+05 | 8228+05 | 1002+06 | 1115+06 |
| 2885+06 | 6288+05 | 6279+05 | 6265+05 | 6236+05 | 6183+05 | 6047+05 | 5964+05 | 5676+05 | 6877+05 | 8636+05 | 1021+06 | 1120+06 |
| 3550+06 | 6977+05 | 6970+05 | 6957+05 | 6933+05 | 6888+05 | 6771+05 | 6686+05 | 6403+05 | 7479+05 | 9114+05 | 1048+06 | 1132+06 |
| 4367+06 | 7741+05 | 7734+05 | 7724+05 | 7703+05 | 7664+05 | 7566+05 | 7483+05 | 7211+05 | 8153+05 | 9660+05 | 1084+06 | 1153+06 |
| 5372+06 | 8588+05 | 8582+05 | 8573+05 | 8555+05 | 8522+05 | 8438+05 | 8362+05 | 8107+05 | 8913+05 | 1027+06 | 1133+06 | 1187+06 |
| 6608+06 | 9527+05 | 9522+05 | 9514+05 | 9498+05 | 9469+05 | 9399+05 | 9330+05 | 9098+05 | 9772+05 | 1098+06 | 1193+06 | 1238+06 |
| 8129+06 | 1057+06 | 1056+06 | 1056+06 | 1054+06 | 1052+06 | 1046+06 | 1040+06 | 1019+06 | 1074+06 | 1181+06 | 1263+06 | 1303+06 |
| 1000+07 | 1172+06 | 1172+06 | 1171+06 | 1170+06 | 1168+06 | 1163+06 | 1157+06 | 1139+06 | 1183+06 | 1277+06 | 1347+06 | 1380+06 |

*Table 8 Hugoniot data* (initial state $\rho_1$=0.171 g/cm$^3$, $T_1$=20.6 K)

| T, K | $\rho$ (g/cm$^3$) | P(Mbar) |
|---|---|---|
| 2000+04 | 8461-00 | 3851-01 |
| 2460+04 | 8676-00 | 4876-01 |
| 3027+04 | 8892-00 | 6168-01 |
| 3723+04 | 9138-00 | 7831-01 |
| 4580+04 | 9447-00 | 1003-00 |
| 5635+04 | 9804-00 | 1297-00 |
| 6931+04 | 1015+01 | 1687-00 |
| 8527+04 | 1039+01 | 2196-00 |
| 1049+05 | 1049+01 | 2858-00 |
| 1290+05 | 1046+01 | 3736-00 |
| 1587+05 | 1038+01 | 4927-00 |
| 1953+05 | 1023+01 | 6515-00 |
| 2402+05 | 9993-00 | 8555-00 |
| 2955+05 | 9687-00 | 1111+01 |
| 3635+05 | 9350-00 | 1426+01 |
| 4472+05 | 9014-00 | 1814+01 |
| 5501+05 | 8714-00 | 2300+01 |
| 6768+05 | 8443-00 | 2908+01 |
| 8326+05 | 8211-00 | 3675+01 |
| 1024+06 | 8012-00 | 4643+01 |
| 1260+06 | 7848-00 | 5867+01 |
| 1550+06 | 7704-00 | 7406+01 |
| 1907+06 | 7578-00 | 9334+01 |
| 2346+06 | 7476-00 | 1175+02 |
| 2885+06 | 7383-00 | 1475+02 |
| 3550+06 | 7302-00 | 1848+02 |
| 4367+06 | 7235-00 | 2308+02 |
| 5372+06 | 7172-00 | 2876+02 |
| 6608+06 | 7118-00 | 3573+02 |
| 8129+06 | 7071-00 | 4431+02 |
| 1000+07 | 7033-00 | 5485+02 |